\documentclass[conference]{IEEEtran}
\IEEEoverridecommandlockouts
\usepackage{hyperref}
\usepackage{mathtools,amssymb}
\usepackage{microtype}
\usepackage[caption=false,font=footnotesize]{subfig}
\usepackage{booktabs} 
\usepackage[capitalize]{cleveref}
\crefformat{equation}{(#2#1#3)}
\crefrangeformat{equation}{(#3#1#4) to~(#5#2#6)}
\crefmultiformat{equation}{(#2#1#3)}%
{ and~(#2#1#3)}{, (#2#1#3)}{ and~(#2#1#3)}
\usepackage{bm}
\usepackage[nolist]{acronym}
\usepackage{siunitx}
\usepackage{pgfplots}
\usepackage{tikzscale}
\usetikzlibrary{shapes}
\usetikzlibrary{external}
\usetikzlibrary{plotmarks,positioning}
\graphicspath{{graphics/}}
\pgfplotsset{compat=1.13}

\DeclareMathOperator{\Langevin}{L}

\DeclareMathOperator{\erf}{erf}

\DeclareMathOperator{\e}{e}

\DeclareSIUnit{\bit}{bit}
\newcommand{\mathsc}[1]{{\normalfont\textsc{#1}}}
\newcommand\average[1]{\overline{#1}}

\newcommand\D[2][]{\frac{\partial^{#1}}{\partial #2^{#1}}}

\newcommand{\hyd}{\mathrm{h}}

\newcommand{\f}{\mathrm{f}}
\newcommand{\m}{\mathrm{m}}
\newcommand{\TX}{\mathsc{tx}}
\newcommand{\RX}{\mathsc{rx}}
\newcommand{\ob}{\mathrm{ob}}

\newcommand{\sat}{\mathrm{s}}
\newcommand{\coat}{\mathrm{c}}
\newcommand{\boltz}{\mathsc{b}}

\newif\ifLongExp
\LongExpfalse
\newif\ifArxiv
\Arxivtrue

\begin{document}
\title{
    Molecular Communication using \\ Magnetic Nanoparticles
}

\author{
    \IEEEauthorblockN{
        Wayan~Wicke,
        Arman~Ahmadzadeh,
        Vahid~Jamali,\\
        and~Robert~Schober
    }
    \IEEEauthorblockA{
        Institute for Digital Communications\\
        Friedrich-Alexander-Universit\"at Erlangen-N\"urnberg
    }
    \and
    \IEEEauthorblockN{
        Harald~Unterweger and Christoph~Alexiou
    }
    \IEEEauthorblockA{
        Section for Experimental Oncology and Nanomedicine\\
        Universit\"atsklinikum Erlangen\\
        Friedrich-Alexander-Universit\"at Erlangen-N\"urnberg
}
    \thanks{%
        This work was supported in part by the Friedrich-Alexander-Universit\"at Erlangen-N\"urnberg (FAU) under the Emerging Fields Initiative (EFI).%
    }
}%
\maketitle%

\begin{abstract}
    In this paper, we propose to use magnetic nanoparticles as information carriers for molecular communication.
    This enables the use of an external magnetic field to guide information-carrying particles towards the receiver.
    We show that the particle movement can be mathematically modeled as diffusion with drift.
    Thereby, we reveal that the key parameters determining the magnetic force are particle size and \emph{magnetic field gradient}.
    As an example, we consider magnetic nanoparticle based communication in a bounded two-dimensional environment. For this model, we derive an analytical expression for the channel impulse response subject to fluid flow and magnetic drift.
    Numerical results, obtained by particle-based simulation, validate the accuracy of the derived analytical expressions.
    Furthermore, adopting the symbol error rate as performance metric, we show that using magnetic nanoparticles facilitates reliable communication, even in the presence of fluid flow.
\end{abstract}


\maketitle
\begin{acronym}
    \acro{ISI}{inter-symbol interference}
    \acro{MC}{Molecular communication}
    \acro{MNP}{magnetic nanoparticle}
    \acro{TX}{transmitter}
    \acro{RX}{receiver}
    \acro{OOK}{on-off keying}
    \acro{ODE}{ordinary differential equation}
    \acro{PDE}{partial differential equation}
    \acro{PDF}{probability density function}
    \acro{SER}{symbol error rate}
\end{acronym}

\section{Introduction}
\ac{MC} is one of the mechanisms that biological cells use to communicate with each other~\cite[Ch.~16]{alberts_essential_2013}.
In natural \ac{MC} systems information is conveyed by specific patterns of molecule releases, e.g., by releasing different numbers or types of molecules.  
Thereby, in typical diffusive \ac{MC} environments, the \emph{information molecules} propagate by \emph{Brownian motion} where the movement of particles is due to thermally induced collisions with molecules of the embedding liquid.

A recent trend in biotechnology is to create artificial and genetically modified cells~\cite[Ch.~10]{alberts_essential_2013}.
These synthetic \emph{nanomachines}, e.g., drug bearing cells, could cooperate and fight a local infection site by adjusting the release of the pharmaceutical in a coordinated and controlled manner~\cite[Ch.~8]{nakano_molecular_2013}.
For this smart collaboration of nanomachines, communication is essential.
Thereby, a message might trigger a certain chemical process which in turn may cause a desired action of a receiving nanomachine.
In this context, \ac{MC} has recently attracted considerable attention as a biocompatible approach for synthetic communication at the cellular level.

For \ac{MC}, usually naturally occurring information molecules such as proteins are considered as information carriers~\cite[Ch.~2]{nakano_molecular_2013}.  
However, apart from problems in realizing synthetic biological \ac{MC} systems at nanoscale~\cite{farsad_comprehensive_2016}, there are also severe limitations by design.
In particular in diffusive \ac{MC}, the movement of the information carriers is random and cannot be stirred towards the receiver, i.e., many of the released molecules do not arrive at the receiver.
Moreover, molecules suspended in a fluid are very sensitive to fluid flow which easily dominates the diffusive movement and in many cases cannot be controlled externally, e.g., in blood vessels.
In this paper, we will show that these problems can be overcome by using \acp{MNP} as information carriers and by guiding them via an external magnetic field.

For targeted drug delivery and many other biotechnological applications, \acp{MNP} are widely used already~\cite{pankhurst_progress_2009,gijs_magnetic_2004}.
These particles usually consist of a polymer matrix with embedded iron oxides which we will simply refer to as \emph{magnetic core}, and a nonmagnetic coating.
The coating ensures biocompatibility and stability, i.e., it prevents agglomeration of the nanoparticles.
Moreover, the particle surface can be \emph{functionalized} with binding sites that are selective to specific molecules~\cite{veiseh_design_2010}.
In this way, \acp{MNP} can be chemically recognized by cells.
Also, by exploiting their magnetic properties, \acp{MNP} can be detected by external devices~\cite{pankhurst_progress_2009}.
However, most importantly, \acp{MNP} can be externally guided by applying a magnetic field.
Thereby, the magnetic force crucially depends on the magnetic field gradient rather than the magnitude of the magnetic field.
Thus, larger forces can be realized by optimizing the design of the magnet to achieve large magnetic field gradients, see e.g.~\cite{sarwar_optimal_2012} where the arrangement of spatial arrays of permanent magnets is optimized for this purpose.

Despite their widespread use in contemporary biotechnology, to the best of our knowledge, for synthetic \ac{MC}, \acp{MNP} have only been considered in~\cite{nakano_externally_2014, kisseleff_magnetic_2017}.
In particular, the authors of~\cite{nakano_externally_2014} proposed that \acp{MNP} attached to DNA can initiate gene expression if subjected to an external magnetic field.
On the other hand, a wearable device detecting changes of inductance when \acp{MNP} pass through a coil was devised in~\cite{kisseleff_magnetic_2017}. 
However, using \acp{MNP} as information carriers and guiding them by an external magnetic field has not been investigated yet.

Motivated by the general availability and applicability of \acp{MNP} in
biotechnology, in this paper,  we make the following contributions:
\begin{enumerate}
    \item
        We propose the use of \acp{MNP} as information carriers and characterize their physical properties.
        Thereby, we model the particle movement in an external magnetic field  as diffusion with drift similar to fluid flow.
        In contrast to fluid flow, a magnetic field can be applied in a desired direction and even towards solid boundaries.
        Moreover, we show that the magnetic force critically depends on the particle size.
    \item
        To illustrate the utility of using \acp{MNP} for \ac{MC}, we consider a generic bounded two-dimensional environment which can be thought of as a simple abstraction of a microfluidic channel or as a rough approximation of a blood vessel. For this model, we analyze the time-variant spatial particle distribution subject to the combined effect of diffusion, fluid flow, and magnetic drift.
        As particles usually differ in size, we also take into account the typical log-normal distribution of the particle radius~\cite{kiss_new_1999} in our mathematical expressions.
    \item
        For the considered model, we calculate the \ac{SER} to evaluate the system performance.
        Thereby, the system is affected by fluid flow which may prevent information-carrying particles from reaching the \ac{RX}.
        We show that applying a magnetic force can drastically reduce the \ac{SER}.
\end{enumerate}

The remainder of this paper is organized as follows.
In \cref{sec:system_model}, we present the system model and the magnetic properties of \acp{MNP}.
Based on this model, we derive the exact particle distribution and the channel impulse response in \cref{sec:system_analysis}.
Simulation results are provided in \cref{sec:numerical_results}.
Finally, \cref{sec:conclusion} concludes the paper.

\section{System Model}
\label{sec:system_model}
\begin{figure}[!t]
    \centering
    \ifArxiv
    \includegraphics{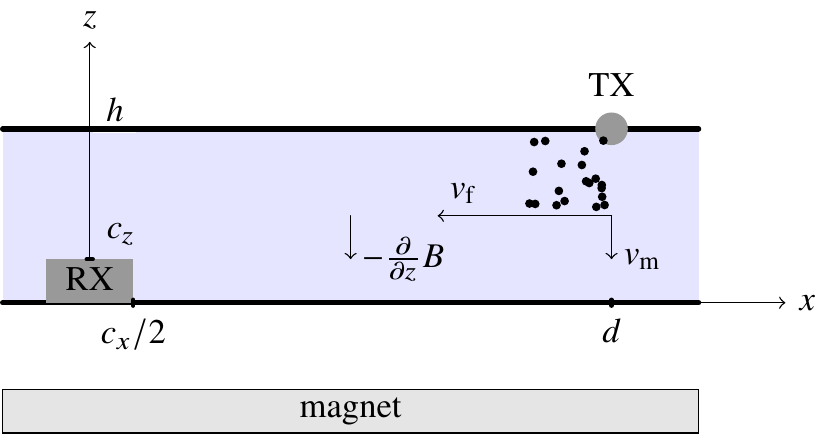}
    \fi
    \caption{\label{fig:system_model}
        System model geometry.
        The overall flow vector and the magnetic field gradient are given by $(-v_\f, -v_\m)$ and $(0,\D{z}B)$, respectively.
    }
\end{figure}

The use of \acp{MNP} can be beneficial for many environments and applications especially when fluid flow hinders \ac{MC}.
For concreteness, as one example where \acp{MNP} can be advantageous, we consider a bounded two-dimensional environment of height $h$ and infinite width in the $x$-$z$-plane. 
In particular, particles can only diffuse within $-\infty<x<\infty, 0\leq z\leq h$ and the boundaries at $z=0$ and $z=h$ are modeled as reflective, see \cref{fig:system_model}.
For this channel, we assume the \ac{TX} positioned at $x=d, z=h$ wants to deliver a message to the \ac{RX} which is located on the opposite side at $x=0, z=0$.
Thereby, fluid flow with velocity $v_\f$ carries the \acp{MNP} downstream in negative $x$-direction but possibly past the \ac{RX}.
To increase the number of particles arriving at the \ac{RX}, a magnet creating a magnetic field $B$ is placed below the channel dragging particles in negative $z$-direction towards the \ac{RX} with velocity $v_\m$.

\subsection{Channel Model}
\label{sec:channelModel}
To illustrate the benefits of using \acp{MNP}, we focus on the following common \ac{MC} model.
The binary information symbols, $b[k]$, are modulated by \ac{OOK}.
Assuming instantaneous particle release, for transmitting $b[k]=1$ and $b[k]=0$, the point source \ac{TX} releases $N_\TX$ and 0 particles, respectively.
The \ac{RX} is a transparent rectangular patch at $|x| \leq c_x/2,\; 0\leq z \leq c_z$ at the bottom of the channel with height $c_z$ and width $c_x$.
We assume that the \ac{RX} is perfectly synchronized with the \ac{TX}, i.e., the \ac{RX} knows the symbol interval $T$ and when transmission starts and ends, see~\cite{jamali_symbol_2017} for more details.
By counting the number of particles within its volume, the \ac{RX} takes samples at times $kT+t_0$, where $t_0$ is the time offset after which, for each symbol interval, $k$, particles can be expected within the receiver volume.
For detection, in each symbol interval, the number of counted particles at the \ac{RX}, $n_\RX[k]$, is compared to a threshold $\xi$, i.e., the detected symbols are given by
\begin{equation}
    \label{eq:bhat}
    \hat{b}[k] = 
    \begin{cases}
        0,	&	n_\RX[k] < \xi \\
        1,	&	n_\RX[k] \geq \xi.
    \end{cases}
\end{equation}

\subsection{Magnetic Nanoparticles}
\label{sec:mnp}
We model the \acp{MNP} by the radius of the magnetic core and the radius including the coating which are denoted as $R_\m$ and $R_\hyd$, respectively.
Because of slight variations in the physical parameters during \ac{MNP} synthesis, actual particle sizes may differ from the intended size.  
Thereby, it has been found that the log-normal distribution usually provides a good fit to the experimentally observed particle sizes~\cite{kiss_new_1999}.
Motivated by this, for the particle radius, $R_\m$, we assume a log-normal distribution with mean $m_R$ and standard deviation $s_R$.
Thereby, for simplicity, we assume that $R_\hyd=R_\m+R_\mathrm{c}$, where $R_\mathrm{c}$ is the constant thickness of the coating.
In the remainder of this paper, we will refer to $m_R$ as the \emph{nominal} particle size, i.e., the target magnetic core size in the production of the \acp{MNP}.

We model the externally applied magnetic field by the magnitude of the magnetic flux density $B$.
Thereby, $B$ is assumed to increase towards the magnet, i.e., the gradient of $B$ is in the negative $z$-direction, cf.~{\cref{fig:system_model}}.
{\acp{MNP}} tend to magnetically align with the applied magnetic field $B$.
However, immersed in a fluid of temperature $T_\f$, the alignment is not perfect.
In particular, considering the thermal energy per particle $k_\mathsc{b}T_\f$, where $k_\textsc{b}\approx\SI{1.381e-23}{\meter^2\kilogram\per\second^2\per\kelvin}$ is the Boltzmann constant~{\cite{nelson_biological_2007}}, the average magnetization in the direction of the magnetic field is given by~{\cite[Ch.~4.3.2]{coey_magnetism_2010}}
\begin{equation}
    \label{eq:hysteresis}
    M(B) = M_\mathrm{s} \Langevin\left(\frac{V_\m 
    M_\mathrm{s}B}{k_\mathsc{b} T_\f}\right),
\end{equation}
where the \emph{Langevin function} $\Langevin(s)$ is defined as $\Langevin(s)=\coth(s) - 1/s$ and $\coth(s)$ is the hyperbolic cotangent. 
Moreover, $V_\m$ is the volume of the magnetic core and $M_\mathrm{s}$ denotes the saturation magnetization which applies if an \ac{MNP} is fully aligned with the applied magnetic field, i.e., $M(B\to\infty)=M_\mathrm{s}$.
In the remainder of this paper, for simplicity, we will assume $M(B)\approx M_\mathrm{s}$.
The validity of this assumption is investigated in \cref{sec:numerical_results}, \cref{fig:hysteresis}.
Eq.~\eqref{eq:hysteresis} implies $M(B=0)=0$, which is in contrast to larger ferromagnetic materials for which $M$ does not only depend on the current value of $B$ but also on previous values of $B$.
This effect is known as \emph{hysteresis}.

Given the volume of the magnetic core $V_\m$ and its average magnetization in \cref{eq:hysteresis}, the force on an \ac{MNP} in a magnetic field $B$ in negative $z$-direction (cf.~\cref{fig:system_model}) is given by~\cite{alexiou_high_2006}
\begin{equation}
    \label{eq:magnetic_force}
    F_\m(z) = -V_\m M_\mathrm{s}\D{z}B(z),
\end{equation}
which is proportional to the magnetic field gradient.
In this paper, we assume that the magnetic field within the channel can be accurately modeled by an affine function of $z$, i.e., we consider the linearization of $B(z)$.
In this case, the magnetic force on the \acp{MNP} is constant.
Thereby, the force points towards the magnet because this is the direction of increasing magnetic field strength.

The movement of the \acp{MNP} is subject to diffusion, which can be characterized by the diffusion coefficient $D$, and a magnetic drift with velocity $v_\m$, which is due to the magnetic force $F_\m$.
It is known that applying a force $F_\m$ on an \ac{MNP} immersed in a liquid of viscosity $\eta$ quickly accelerates it to the terminal velocity $v_\m=F_\m/\zeta$~\cite[Eq.~(4.12)]{nelson_biological_2007}, where $\zeta$ is the friction coefficient which by Stokes' law is given by $\zeta=6\pi\eta R_\hyd$.
In summary, we obtain
\begin{equation}
    \label{eq:myF}
    v_\m = -\frac{2 M_\sat}{9\eta} \frac{R_\m^3}{R_\m + R_\coat} 
    \D{z}B(z),
\end{equation}
which is proportional to the magnetic field gradient and strongly depends on the particle size.

By thermodynamic reasoning~\cite[Ch.~4]{nelson_biological_2007}, $\zeta$ is linked to the diffusion coefficient $D$ by the Einstein relation $k_\textsc{b}T_\mathrm{f} = D\zeta$.
Hence, given the viscosity $\eta$ and the temperature of the fluid $T_\f$, $D$ can be determined as
\begin{equation}
    \label{eq:myD}
    D = \frac{k_\boltz T_\f}{6\pi\eta (R_\m + R_\coat)},
\end{equation}
which also depends on the particle size but to a lesser degree than $v_\m$.
In contrast, the fluid flow velocity $v_\f$ is not affected by the value of $R_\m$.

\section{Performance Analysis}
\label{sec:system_analysis}
In this section, we derive an analytical expression for the time-variant spatial \ac{MNP} distribution by solving the diffusion equation with drift for the system in \cref{fig:system_model}.
Then, equipped with the solution to the diffusion equation, we calculate the probability of observing a particle within the \ac{RX} volume as well as the expected received number of particles which is a function of time that we will refer to as \emph{impulse response}.
Finally, given the impulse response, we determine the average received signal and the \ac{SER}.

\subsection{Impulse Response}
In the environment depicted in \cref{fig:system_model}, the particle movement in the $x$- and $z$-direction is uncoupled and hence the time-varying \ac{PDF} for the \ac{MNP} position can be written as $p(x,z;t) = p_x(x;t) p_z(z;t)$.
In this equation, the horizontal distribution $p_x(x;t)$ corresponds to an unbounded environment with constant drift $v_\f$.
Hence, this distribution is readily obtained as~\cite[Eq.~(4.39)]{schulten_lectures_2000}
\begin{equation}
    \label{eq:px}
    \ifLongExp
    p_x(x;t) = \frac{1}{\sqrt{4D\pi t}} \exp\left(\frac{-(x-d+v_\f 
        t)^2}{4Dt}\right),
    \else
    p_x(x;t) = \frac{1}{\sqrt{4D\pi t}} \e^{-(x-d+v_\f 
        t)^2/(4Dt)},
    \fi
\end{equation}
where the mean particle $x$-coordinate arrives at the \ac{RX} at time $t_1=d/v_\f$.
Determining the vertical distribution $p_z(z;t)$ is more challenging because of the combination of a bounded environment and particle drift.  
Therefore, we consider the underlying \ac{PDE} which is the diffusion equation with drift.
Thereby, the reflective boundary conditions are specified by~\cite[Eq.~(4.24)]{schulten_lectures_2000} $D\D{z} p_z(z;t) + v_\m p_z(z;t)=0$ for $z=0,h$, and $t>0$.
Moreover, by assumption of a point source \ac{TX}, the initial position $z_0=h$ is known a priori.
Hence, $p_z(z;t)$ for $t>0$ is obtained by solving the following \ac{PDE} with boundary and initial conditions
\begin{equation}
    \label{eq:pde_bounded}
    \begin{aligned}
        \D{t} p_z(z;t) &= v_\m\D{z} p_z(z;t) + D\D[2]{z} p_z(z;t), & &0<z<h \\
        \frac{\partial}{\partial z} p_z(z;t)     &= -\frac{v_\m}{D}p_z(z;t), & 
        &z=0, h \\
        p_z(z;t) &= \delta(z-z_0), & &t = 0.
    \end{aligned}
\end{equation}
Solutions to the one-dimensional diffusion equation without drift are well known for various boundary conditions~\cite{carslaw_conduction_1986}.  
Motivated by this, using a variable substitution, we obtain an equivalent problem formulation in terms of an auxiliary function $q(z;t)$ without drift term but with $q(z;t\to\infty)=0$ and modified boundary conditions~\cite{perez_guerrero_analytical_2009}.
To this end, we define $q(z;t)$ by
\begin{equation}
    \label{eq:p_transform}
    \ifLongExp
    p_z(z;t) = q(z;t)\exp\left(-\frac{v_\m}{2D}(z - z_0) - 
    \frac{v_\m^2}{4D}t\right) + \overline{p}_z(z).
    \else
    p_z(z;t) = q(z;t)\e^{-u(z - z_0) - 
    D u^2 t} + p^\mathrm{eq}_z(z),
    \fi
\end{equation}
where $p_z^\mathrm{eq}(z)=p_z(z;t\to\infty)$ is the steady state or \emph{equilibrium} solution of \cref{eq:pde_bounded} and $u=v_\m/(2D)$.
Substituting \cref{eq:p_transform} in \cref{eq:pde_bounded}, for $t>0$ we obtain the following \ac{PDE} in $q(z;t)$
\begin{equation}
    \label{eq:auxiliary_bounded}
    \begin{aligned}
        \D{t} q(z;t) &= D\D[2]{z} q(z;t), & &0<z<h \\
        \frac{\partial}{\partial z} q(z;t)     &= -uq(z;t), & &z=0,h \\
        q(z;t) &= \delta(z-z_0) - p^\mathrm{eq}_z(z) \e^{u(z - z_0)}, & &t 
        = 0,
    \end{aligned}
\end{equation}
which is the diffusion equation without drift and can be solved by separation of variables.
Nevertheless, to obtain $p_z(z;t)$ in \cref{eq:p_transform}, we require $p_z^\mathrm{eq}(z)$.
Therefore, in the following, we first give the steady state solution and then use it to solve the original problem.

\subsubsection{Steady State}
The steady state solution of $p_z(z;t)$ also needs to satisfy \cref{eq:pde_bounded} but is characterized by $\frac{\partial}{\partial t} p_z(z;t)=0$, i.e., the following \ac{ODE}
\begin{equation}
    \label{eq:equilibrium_ode}
    \begin{aligned}
        -\frac{v_\m}{D} p^\mathrm{eq}_z(z)  &= \D{z} p^\mathrm{eq}_z(z), & 
        z&\in [0, 
        h]
    \end{aligned}
\end{equation}
has to be solved.
As $p^\mathrm{eq}_z(z)$ is a \ac{PDF} and no particles are lost, the reflective boundary conditions are met if the steady state \ac{PDF} satisfies
%
    $\int_0^h p^\mathrm{eq}_z(z)\,\mathrm dz = 1.$

As can be verified by substitution, \cref{eq:equilibrium_ode} is solved by
\begin{equation}
    \label{eq:equilibrium_constant}
    \ifLongExp
    p^\mathrm{eq}_z(z) = s^{-1} \exp(-v_\m/D z),
    \else
    p^\mathrm{eq}_z(z) = s^{-1} \e^{-v_\m z/D},
    \fi
\end{equation}
where $s = D\left(1 - \e^{-v_\m h/D}\right)/v_\m$.

\subsubsection{Transient Solution\label{sec:exact_constant_bounded}}
Using separation of variables, we obtain the auxiliary function $q(z;t)$ in terms of the following series solution
\begin{equation}
\label{eq:qSeries}
    q(z;t) = \sum_{n=1}^\infty Z_n(z) \e^{-D s_n^2 t} a_n,
\end{equation}
where $s_n = n\pi/h$,
\begin{equation}
    \label{eq:Zn}
    Z_n(z) = \sqrt{\frac{2}{h(s_n^2 + u^2)}}
    \left(s_n\cos(s_n z) - u\sin(s_n z)\right),
\end{equation}
and $a_n = Z_n(z_0)$.

We are now ready to determine $p_z(z;t)$ by substituting \cref{eq:qSeries,eq:equilibrium_constant} in \cref{eq:p_transform}.
In summary, we obtain
\begin{equation}
\label{eq:constant_bounded_exact_pdf}
\ifLongExp
    \begin{aligned}
    p_z(z; t) = &\exp\left(-\frac{v_\m}{2D}(z-z_0)\right)\sum_{n=1}^\infty
    \exp\left(-\left(Ds_n^2 + 
    \frac{v_\m^2}{4D}\right)t\right)Z_n(z)Z_n(z_0) \\
    &+\frac{v_\m/D}{1 - \exp(-v_\m/D h)} 
    \exp\left(-\frac{v_\m}{D}z\right),
    \end{aligned}
\else
    \begin{aligned}
    p_z(z; t) = p^\mathrm{eq}_z(z) + &\e^{-u(z-z_0)}\times \\
    &\sum_{n=1}^\infty
    \e^{-D(s_n^2 + 
        u^2)t}Z_n(z)Z_n(z_0),    
    \end{aligned}
\fi
\end{equation}
which simplifies to \cref{eq:equilibrium_constant} for $t\to\infty$. 

\subsubsection{Probability of Particle Observation}
Using the \ac{PDF} $p(x,z;t)$, we can now determine the probability of observing a particle within the \ac{RX} volume, $P_\ob(t)$, as
\begin{equation}
\label{eq:received_in_tube}
P_\ob(t) = P_{\ob,x}(t) P_{\ob,z}(t),
\end{equation}
where $P_{\ob,x}(t)$ and $P_{\ob,z}(t)$ are the probabilities of observing a particle within the \ac{RX} $x$- and $z$-coordinates $[-c_x/2,c_x/2]$ and $[0, c_z]$, respectively.
Integrating \cref{eq:px} from $-c_x/2$ to $c_x/2$ yields
\begin{equation}
    \label{eq:one_dimension_observation_probability}
    P_{\ob,x}(t) = \frac{1}{2} \left[\erf\left(\frac{\overline{x}(t) +
                \frac{1}{2}c_x}{\sqrt{4Dt}}\right) - \erf\left(\frac{\overline{x}(t) -
    \frac{1}{2}c_x}{\sqrt{4Dt}}\right)\right],
\end{equation}
where $\erf(s)$ is the \emph{error function} and $\overline{x}(t)=d-v_\f t$.
Furthermore, integrating \cref{eq:constant_bounded_exact_pdf} from $0$ to $c_z$ yields $P_{\ob,z}(t)$ as
\begin{equation}
\label{eq:constant_bounded_exact}
    \ifLongExp
    \begin{aligned}
        P_{\ob,z}(t) = 
        &\exp\left(-\frac{v_\m}{2D}(c_z-z_0)\right)\sum_{n=1}^\infty
        \exp\left(-\left(Ds_n^2 + 
        \frac{v_\m^2}{4D}\right)t\right)L_n(z)Z_n(z_0) \\
        &+\frac{1 - \exp(-v_\m c_z/D)}{1 - \exp(-v_\m h/D)},
    \end{aligned}
    \else
    \begin{aligned}
        P_{\ob,z}(t) = P^\mathrm{eq}_{\ob,z} +
        &\e^{-u(c_z-z_0)} \times\\
        &\sum_{n=1}^\infty
        \e^{-D(s_n^2 + 
        u^2)t}L_n Z_n(z_0),\\
    \end{aligned}
    \fi
\end{equation}
where $L_n$ is obtained as
\begin{equation}
    L_n = \sqrt{\frac{2}{h(s_n^2 + u^2)}} \sin(s_n c_z),
\end{equation}
and $P^\mathrm{eq}_{\ob,z}$ is the integral of \cref{eq:equilibrium_constant} which is easily found as
\begin{equation}
    \label{eq:myPeq}
    P^\mathrm{eq}_{\ob,z} = \frac{1 - \e^{-v_\m c_z/D}}{1 - \e^{-v_\m h/D}}.
\end{equation}

We note that $P_{\ob,z}(t)$ in \cref{eq:constant_bounded_exact} is the sum of a transient term approaching zero for $t\to\infty$ and a constant steady state term.
Therefore, we can make the following \emph{equilibrium approximation}.
If transmitter and receiver are placed far apart, at time $t_1$ when the particles are expected at the $x$-coordinates of the receiver, $p_z(z;t)$ will have converged to $p^\mathrm{eq}_z(z)$.
Consequently, in this case, the particle observation probability reduces to
%
$P_\ob(t) = P_{\ob,x}(t) P^\mathrm{eq}_{\ob,z}$,
%
which is a simple scaling of $P_{\ob,x}(t)$ in \cref{eq:one_dimension_observation_probability}.
By considering $P^\mathrm{eq}_{\ob,z}$ in \cref{eq:myPeq}, we can also gain qualitative insight on how the particle distribution is affected by changes of the magnetic drift velocity $v_\m$.
In particular, for $v_\m\to 0$, we obtain $P^\mathrm{eq}_{\ob,z}=c_z/h$ as in this case the particle distribution is uniform across the channel height because of diffusion.
On the other hand, for $v_\m\to \infty$, we obtain $P^\mathrm{eq}_{\ob,z}=1$, i.e., all particles are gathered at the lower boundary in \cref{fig:system_model} since the magnetic drift completely dominates diffusion.

Having obtained the particle observation probability at the receiver \cref{eq:received_in_tube}, we can now obtain the expected number of received molecules as a function of time due to the release of $N_\TX$ particles.
Thereby, assuming the number of particles within the channel is small enough such that particle interactions can be ignored, we obtain the \emph{impulse response}
\begin{equation}
    \label{eq:myNob}
    \average{N}_\ob(t) = \sum_{i=1}^{N_\TX} P_{\ob,i}(t),
\end{equation}
where $P_{\ob,i}(t)$ is the probability of observing particle $i$ at time $t$.
Thereby, the $P_{\ob,i}(t),\; i=1,2,\dots,N_\TX$, vary for $s_R>0$ as the particle sizes differ.
In particular, in \cref{eq:one_dimension_observation_probability,eq:constant_bounded_exact}, $v_\m$ and $D$ depend on $R_\m$ via \cref{eq:myF,eq:myD}.
On the other hand, for $s_R=0$, we have $P_{\ob,i}(t)=P_{\ob}(t),\;\forall i$, i.e., all particles have the same (nominal) particle size $m_R$.  
In this case, $\average{N}_\ob(t) = N_\TX P_{\ob}(t)$ which we will refer to as the \emph{nominal impulse response}.

\subsection{Symbol Error Rate}
Using~\cite[Eq.~(30)]{noel_improving_2014}, the average number of observed particles $\average{n}_\RX[k]$ in the $k$-th time slot due to a binary sequence of transmitted symbols $b[k]\in\{0, 1\},\; 0\leq k<K$, is given by
\begin{equation}
\average{n}_\RX[k] = \sum_{i=0}^{k} b[i]\overline{N}_\ob((k-i)T
+ t_0).
\end{equation}

In general, using the detection rule in \cref{eq:bhat}, the probability of making an error in the $k$-th symbol can be written as
\begin{equation}
    \label{eq:myPe}
    \Pr(\hat{b}[k]\neq b[k];\; 
    b[\kappa\leq k]) =
    \begin{cases}
        p_\xi, & b[k] = 1 \\
        1-p_\xi, & b[k] = 0,
    \end{cases}
\end{equation}
where $p_\xi=\Pr(n_\RX[k]< \xi;\; b[\kappa\leq k])$ is the probability of observing less than $\xi$ \acp{MNP} at the $k$-th sampling time given $b[\kappa]$ for $0\leq\kappa\leq k$.
Similar to \cite{noel_improving_2014}, $n_\RX[k]$ can be well approximated by a Poisson random variable with mean $\average{n}_\RX[k]$, see~\cite{le_cam_approximation_1960}.
In this case, $p_\xi=\Pr(n_\RX[k]\leq \xi-1;\; b[\kappa\leq k])$ is the Poisson \emph{cumulative distribution function} with mean $\average{n}_\RX[k]$ evaluated at $\xi-1$.

If there is no \ac{ISI}, then for any $\xi\geq 1$, $b[k]=0$ is always detected correctly because in this case $\average{n}_\RX[k]=0$.
On the other hand, if $b[k]=1$, then $\average{n}_\RX[k]=\average{N}_\ob(t_0)$ and the error probability is minimized for the minimal $\xi$.
Hence, assuming there is no \ac{ISI}, $\xi=1$ minimizes $P_\mathrm{e}$.
In this case, an error occurs only if $b[k]=1$ and $n_\RX[k]=0$.
Hence, the average symbol error rate is given by $P_\mathrm{e}=1/2\times\Pr(n_\RX[k]=0;\; b[k]=1)$ assuming $\Pr(b[k]=1)=1/2$.
For Poisson random variable $n_\RX[k]$, the average \ac{SER} simplifies to
\begin{equation}
    \label{eq:myPeApprox}
    P_\mathrm{e} = \frac{1}{2} \e^{-\overline{N}_\ob(t_0)},
\end{equation}
which we will refer to as the \emph{no \ac{ISI} approximation}.

\section{Numerical Results}
\label{sec:numerical_results}
In this section, unless explicitly stated otherwise, we adopt the physical parameters in \cref{tab:parameters} using the viscosity of water for $\eta$, room temperature for $T_\f$, a saturation magnetization similar to magnetite for $M_\mathrm{s}$, a magnetic field gradient well within the range of realizable values for $\left|\D{z}B\right|$~\cite{alexiou_high_2006}, and values for $h$ and $v_\f$ realizable in a microfluidic setting~\cite{gijs_magnetic_2004}.
Thereby, we choose $t_0=t_1$, where $t_1=d/v_\f$, i.e., the \ac{RX} takes a sample when particles are expected due to fluid flow.
\begin{table}
    \caption{
        \label{tab:parameters}
        System Parameters.
    }
    \centering
        \begin{tabular}{llll}
            \toprule
            Parameter & Value & Parameter & Value \\
            \cmidrule(r){1-2} \cmidrule(r){3-4}
            $\eta$ & \SI{1e-3}{\kilogram\per\meter\per\second} & $d$ & \SI{1}{\milli\meter} \\
            $T_\f$ & \SI{300}{\kelvin} & $h$ & \SI{10}{\micro\meter}\\
            $R_\mathrm{c}$ & \SI{1}{\nano\meter} & $c_x$ & \SI{0.1}{\milli\meter}\\
            $m_R$ & \SI{50}{\nano\meter} & $c_z$ & \SI{1}{\micro\meter}\\
            $s_R$ & \SI{10}{\nano\meter} & $v_\f$ & \SI{0.5}{\milli\meter\per\second} \\
            $M_\mathrm{s}$ & \SI{5e5}{\ampere\per\meter} & $T$ & \SI{2}{\second} \\
            $\left|\D{z}B\right|$ & \SI{5}{\tesla\per\meter} & $\xi$ & $1$ \\
            \bottomrule
        \end{tabular}
\end{table}
For simulation of the system described in \cref{sec:system_model}, we use a particle-based approach where time advances in discrete time steps $\Delta t$ and the position of each particle is tracked and updated in each time step, see e.g.~\cite[Eq.~(1)]{farsad_comprehensive_2016}.
Then, for the received signal, for each time step, the number of particles within the receiver volume is counted.
Within each simulation step, if a particle crosses a channel boundary, it is reflected back into the channel.

\begin{figure}[!t]
    \ifArxiv
    \includegraphics{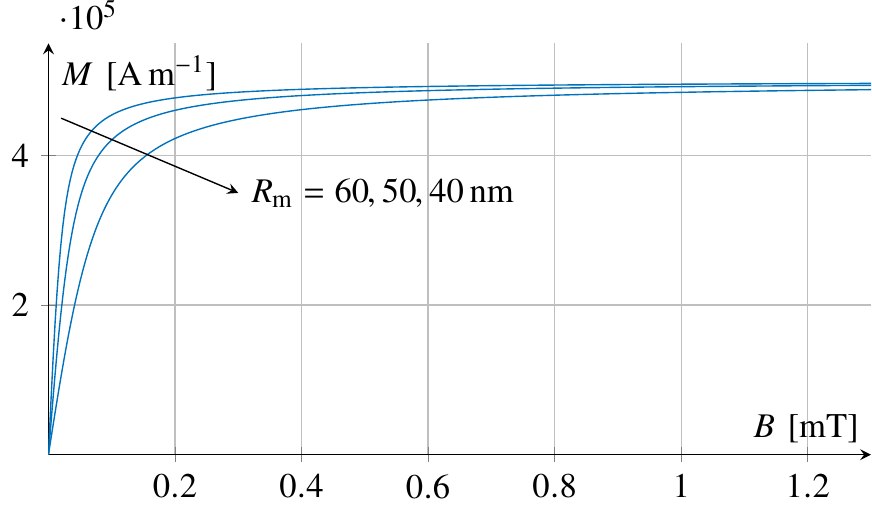}
    \fi
    \caption{\label{fig:hysteresis}
        Magnetization curves as given by \cref{eq:hysteresis} for magnetic core sizes $R_\m=40, 50, \SI{60}{\nano\meter}$.
    }
\end{figure}

In \cref{fig:hysteresis}, we evaluate the particle magnetization \cref{eq:hysteresis} as a function of the applied magnetic field $B$ for $R_m=m_R=\SI{50}{\nano\meter}$ and $R_m=m_R\pm s_R$.
As $M(B)$ in \cref{eq:hysteresis} is point symmetric, we only show $M(B)$ for $B\geq 0$.
Here, magnetization is saturated already for $B\approx \SI{1}{\milli\tesla}$ which is easily exceeded by today's magnets for targeted drug delivery~\cite{alexiou_high_2006}.
Hence, assuming $M(B)\approx M_\mathrm{s}$ in \cref{sec:system_model}, is justified.
Compared to the nominal particle size, slightly larger and smaller \acp{MNP} reach saturation quicker and slower because their magnetization direction is less and more affected by thermal energy, respectively.

%

\begin{figure}[!t]
    \ifArxiv
    \includegraphics{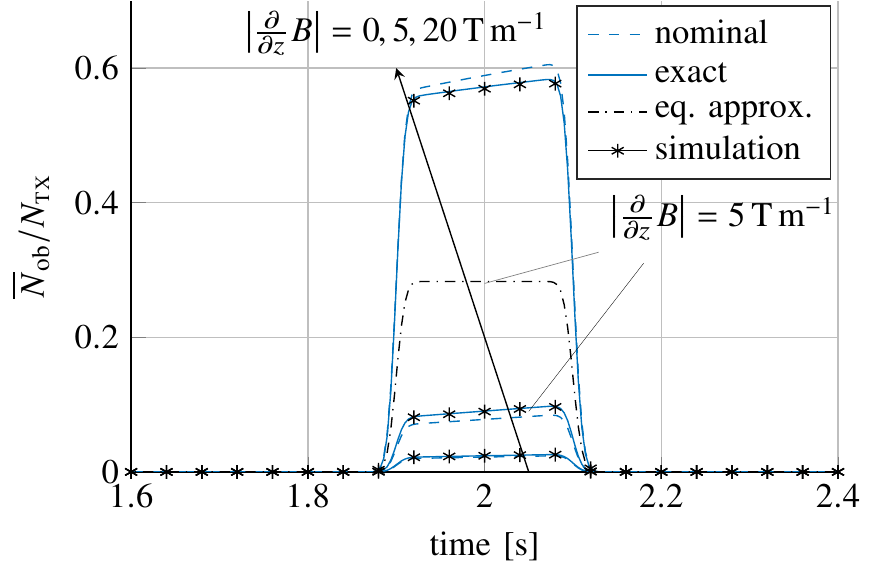}
    \fi
    \caption{
        \label{fig:ir}
        Impulse response $\average{N}_\ob(t)$ in \cref{eq:myNob} for different magnetic field gradients where $N_\TX=1000$.
        Simulation results with $\Delta t=\SI{2}{\milli\second}$ have been averaged over $10^4$ realizations (for clarity, not all points are shown).
    }
\end{figure}

In \cref{fig:ir}, we show the fraction of \acp{MNP} within the receiver volume after a point release of $N_\TX$ particles at the transmitter at $t=0$ as a function of time for times around $t_1=\SI{2}{\second}$.
The curves are parameterized by different magnetic field gradients $\left|\D{z}B\right|$ resulting in different drift velocities $v_\m$.
For each $\left|\D{z}B\right|$, we also show the nominal impulse response as well as simulated data points.
For clarity, the equilibrium approximation is only shown for $\left|\D{z}B\right|=\SI{5}{\tesla\per\meter}$.
\cref{fig:ir} shows that increasing the magnetic field gradient significantly increases the number of observed \acp{MNP}.
There is a time window of approximately $\SI{0.2}{\second}$ centered around $t_1=\SI{2}{\second}$ within which a nonzero number of \acp{MNP} can be observed independent of the magnetic field gradient.
Hence, for a symbol interval size of $T=\SI{2}{\second}$, \ac{ISI} does not play a significant role for the given parameters due to the flow-dominated transport of particles.
Due to the log-normal particle size distribution there is some deviation from the nominal impulse response as $v_\m$ depends on the particle size.
In particular, when $\left|\D{z}B\right|$ is relatively small (e.g. $\left|\D{z}B\right| = \SI{5}{\tesla\per\meter}$) and large  (e.g. $\left|\D{z}B\right| = \SI{20}{\tesla\per\meter}$), more and fewer \acp{MNP} are observed at the \ac{RX} than expected based on the nominal impulse response, respectively.
This can be explained as follows.
The nominal impulse response is obtained based on the assumption that the radius of all \acp{MNP} is equal to the mean radius $m_R$.
The magnetic force experienced by a particle increases with its size.
Hence, for small $\left|\D{z}B\right|$, the magnetic force experienced by \acp{MNP} having radius $m_R$ is relatively weak and having \acp{MNP} with a larger radius, and thus, experiencing  a larger magnetic force, increases the number of observed \acp{MNP}.
On the other hand, for large  $\left|\D{z}B\right|$, the magnetic force experienced by \acp{MNP} having radius $m_R$ is relatively strong and almost all \acp{MNP} with this radius arrive at the \ac{RX}.
Thus, having even larger \acp{MNP} cannot further increase the number of observed \acp{MNP}.
However, having smaller \acp{MNP}, which experience a weaker magnetic force, decreases the number of observed \acp{MNP}.
Regarding the equilibrium approximation for $\left|\D{z}B\right|=\SI{5}{\tesla\per\meter}$, we see that it overestimates the number of observed particles as within the time frame where \acp{MNP} can be observed, the steady state has not yet been reached.
Overall, \cref{fig:ir} confirms that magnetically targeting the \ac{RX} proves effective in increasing the number of observed \acp{MNP}.
Thereby, for the considered system model, larger magnetic field gradients are preferable.

\begin{figure}[!t]
    \ifArxiv
    \includegraphics{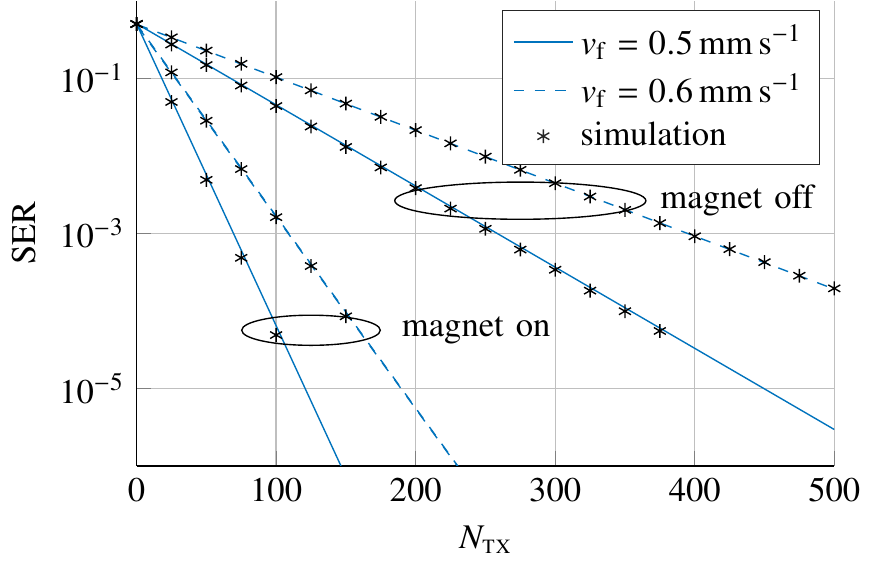}
    \fi
    
    \caption{\label{fig:ser}
        Symbol error rate as a function of the available number of \acp{MNP} per symbol.
        $P_\mathrm{e}$ in \cref{eq:myPeApprox} with $K=10$ is shown for two different fluid flow velocities and with the magnet turned on and off, respectively. 
        Simulation results with $\Delta t=\SI{20}{\milli\second}$ have been averaged over $10^6$ realizations.
    }
\end{figure}

In \cref{fig:ser}, we evaluate the symbol error rate when the magnet is turned on and off, respectively, as a function of the number of \acp{MNP} used per transmit pulse.
In particular, for each $N_\TX$, we show the \ac{SER} according to \cref{eq:myPeApprox} for $\xi=1$ as well as simulation results for $\mathrm{SER}\geq \SI{5e-5}{}$ where the particle sizes have been chosen independently for each transmit pulse of $N_\TX$ particles.
As for $T=\SI{2}{\second}$ no \ac{ISI} is expected for the chosen system parameters, cf.~\cref{fig:ir}, the no \ac{ISI} approximation in \cref{eq:myPeApprox} matches the simulation results.
We can also observe that the system is very sensitive to changes in the fluid flow which can not be controlled externally.
However, overall, we note that turning the magnet on reduces the \ac{SER} significantly.

\section{Conclusion}
\label{sec:conclusion}
In this paper, we proposed the use of \acp{MNP} as information-carriers for \ac{MC} systems.
In particular, we showed how the movement of \acp{MNP} can be modeled as diffusion with drift.
To this end, we reviewed the magnetic drift velocity resulting from a magnetic force caused by a magnetic field gradient.
\newpage
Thereby, we highlighted the dependence of the drift velocity and the diffusion coefficient on the particle size.
Subsequently, we introduced a technique to solve the diffusion equation with drift in a bounded environment and applied this technique to derive the impulse response in a two-dimensional environment subject to fluid flow, diffusion, and magnetic drift.
Moreover, we showed how the particle size distribution can be incorporated in the impulse response.
By numerical evaluation, we illustrated how a log-normal particle size distribution affects the impulse response for different magnetic field gradients.
Finally, evaluating the \ac{SER} revealed the sensitivity of the system performance to variations in the fluid flow and demonstrated the effectiveness of employing external magnetic fields for improving the reliability of communication.

\bibliographystyle{IEEEtran}
\bibliography{main} 

\end{document}